\documentclass[aps,amsmath,amssymb, reprint, prl]{revtex4-1}

\usepackage{hyperref}
\usepackage{xcolor}
\usepackage{graphicx}
\usepackage{amsfonts,amssymb,amsmath}

\definecolor{darkblue}{rgb}{0.1,0.1,0.7}
\definecolor{darkred}{rgb}{0.5,0.1,0.1}
\definecolor{darkgreen}{rgb}{0.0,0.42,0.06}
\hypersetup{colorlinks=true,urlcolor=darkred,linkcolor=darkblue,citecolor=darkblue}
\definecolor{shadecolor}{rgb}{0.85,0.85,0.85}

\newcommand{\bC}{\ensuremath{\mathbb{C}}}

\newcommand{\bZ}{\ensuremath{\mathbb{Z}}}

\newcommand{\cA}{\mathcal{A}}
\newcommand{\cB}{\mathcal{B}}

\newcommand{\cH}{\mathcal{H}}

\newcommand{\cN}{\mathcal{N}}

\newcommand{\cR}{\mathcal{R}}

\newcommand{\SU}{\mathrm{SU}}

\newcommand{\SL}{\mathrm{SL}}

\newcommand{\U}{\mathrm{U}}

\newcommand{\Tr}{\operatorname{Tr}}
\newcommand{\PF}{\operatorname{PF}}
\newcommand{\MM}{\operatorname{MM}}

\def \wt  {\widetilde}

\def \be  {\begin{equation}}
\def \ee  {\end{equation}}
\def \bea {\begin{equation}\begin{aligned}}
\def \eea {\end{aligned}\end{equation}}

\begin{document}

\title{New conformal field theory from $\cN=(0,2)$ Landau-Ginzburg model}

\author{Jirui Guo${}^1$, Satoshi Nawata${}^1$, Runkai Tao${}^{1,2}$, Hao Derrick Zhang${}^1$}
\affiliation{${}^1$Department of Physics and Center for Field Theory and Particle Physics, Fudan University, 220
Handan Road, 200433 Shanghai, China}\email{snawata@gmail.com}
\affiliation{${}^2$Department of Physics and Astronomy, Rutgers University, 126 Frelinghuysen Road, Piscataway NJ 08854, USA}

\begin{abstract}
By studying the infra-red fixed point of an $\cN=(0,2)$ Landau-Ginzburg model, we find an example of modular invariant partition function beyond the ADE classification. This stems from the fact that a part of the left-moving sector is a new conformal field theory which is a variant of the parafermion model.

\vspace{.5cm}
\begin{center}
\emph{Dedicated to the academic achievements of Tohru Eguchi and Sung-Kil Yang}\end{center}
\end{abstract}

\setcounter{tocdepth}{2}
\maketitle

\section{Introduction}
A 2d CFT is endowed with an infinite-dimensional Lie algebra \cite{belavin1984infinite}, and modular invariance further constrains its spectrum on the torus \cite{cardy1986operator}. Consequently, a number of models have been exactly solved. (For instance, see \cite{itzykson1988conformal}.)
In an RCFT, a modular invariant partition function consists of finitely many pairs $I$ of left- and right-moving characters of chiral algebras $\cA\otimes \overline \cA$
\be\label{RCFT-Hilbert}
Z=\sum_{(i,\overline i)\in I}N_{i\overline i} ~\chi_{i}^\cA \otimes \overline \chi_{\overline i}^{\overline\cA}~.
\ee
If we write $M_{ij}^\cA~\rotatebox[origin=c]{-90}{$\circlearrowright$}~\{\chi_{j}\}$ and $(M_{\overline i\overline j}^{\overline \cA})^*~\rotatebox[origin=c]{-90}{$\circlearrowright$}~\{\overline \chi_{\overline j}\}$ for actions of $M\in \SL(2,\bZ)$ on the spaces of the left and right-moving characters \cite{Kac:1984mq}, then the modular invariance requires
$$M_{ij}^\cA N_{j\overline j}(M_{\overline j\overline i}^{\overline \cA})^*=N_{i\overline i}~.$$
 As a result, modular invariant partition functions of $\SU(2)_k$ WZNW models and unitary Virasoro minimal models admit the celebrated
ADE classifications \cite{cappelli1987modular,cappelli1987ade,kato1987classification,Gannon:1999cp}. If a CFT is described by a non-chiral coset model \cite{goddard1986unitary} involving $\SU(2)$ and $\U(1)$, its modular invariant partition function fits into the ADE classification. As such,
one can find modular invariant partition functions for parafermion (PF) models $\SU(2)_k/\U(1)_k$ \cite{Gepner:1986hr} and $\cN=2$ minimal models (MMs) $(\SU(2)_k\times \U(1)_2)/\U(1)_{k+2}$ \cite{Ravanini:1987yg}. Furthermore, with $\cN=(2,2)$ supersymmetry, Landau-Ginzburg (LG) models with ADE quasi-homogeneous superpotential are described by the MMs of corresponding ADE type in the infra-red (IR) limit \cite{Vafa:1988uu,Martinec:1988zu}.

On the other hand, the class of $\cN=(0,2)$ LG models is much richer because firstly they are chiral in general and secondly there is more freedom due to the $E$- and $J$-terms \cite[\S6]{Witten:1993yc}. Therefore, it is natural to ask how IR CFTs incorporate the richness of $\cN=(0,2)$ LG models by encoding the information of the $E$- and $J$-terms.

In this article, we make a modest step towards understanding the LG/CFT correspondence with $\cN=(0,2)$ supersymmetry  by studying the IR fixed point of a certain $\cN=(0,2)$ LG model along the line of \cite{Gadde:2016khg}. We will obtain its modular invariant partition function, which turns out to be beyond the ADE classification. Careful analysis of the Hilbert space will show that a part of the left-moving sector is described by a new CFT which is a close cousin of the parafermion model.

\section{LG/CFT correspondence}
To begin with, we describe the $\cN=(0,2)$ LG model we focus on. It is a theory of two chiral multiplets $\phi_1,\phi_2$ and two Fermi multiplets $\psi_1,\psi_2$ with interactions determined by a superpotential
\be\label{potential}
W = \psi_1 (\phi_1^4+\phi_2^2) + \psi_2 \phi_1^2\phi_2~.
\ee
The $E$-term is set to zero.
This theory is called Class 2.b with $k=4$ in \cite{Gholson:2018zrl}.

Since the numbers of chiral and Fermi multiplets are equal, the vanishing of the gravitational anomaly $\Tr \gamma_3=\overline{c}-c=0$ guarantees the equality of the left- and right-moving central charges. Furthermore, the $c$-extremization \cite{Benini:2012cz} calculates
\be\label{central-charge}
c=\overline{c} = \frac{75}{27}~,
\ee
where the $\cR$-charges of all the multiplets are listed in the following table.
\begin{table}[h]
\begin{tabular}{c|cccc}
& $\phi_1$ & $\phi_2$ & $\psi_1$ & $\psi_2$ \\\hline
$\U(1)_\cR$ & $\frac{5}{27}$ & $\frac{10}{27}$ & $\frac{7}{27}$ & $\frac{7}{27}$ \\
$\U(1)_\ell$ & $1$ & $2$ & $-4$ & $-4$
\end{tabular}
\end{table}
There is also a left-moving $\U(1)_\ell$ global symmetry
with 't Hooft anomaly $27$.  Therefore, these data suggest that, in the IR fixed point, the right-moving sector is the $\cN=2$ MM${}_{25}$ with level $k=25$, and the left-moving sector is the $\U(1)_{\frac{27}2}$ WZNW model with level $k=27/2$ and a CFT of central charge $16/9$. It is tempting to identify the CFT of central charge $16/9$ with the parafermion model $\PF_{25}$ of level $25$ as in \cite{Gadde:2016khg}, and we will indeed write a modular invariant partition function using characters of $\PF_{25}$ in the next section. However, as we will see later, it is not exactly the $\PF_{25}$, but a certain variant of the $\PF_{25}$.

Let us extract more information about the IR CFT from the UV data.
Since an elliptic genus is protected under the RG flow \cite{Dedushenko:2015opz}, it can be computed from the information of the LG model. We evaluate it in the NS-NS sector
\bea\label{EG}
\mathrm{EG}(\tau,z)&= \Tr_{\textrm{NSNS}}(-1)^{F} q^{L_{0}-\frac{c}{24}} y^{J_{0}} e^{-\beta\left(\overline{L}_{0}-\frac{1}{2} \overline{J}_{0}\right)}\cr
&=q^{-\frac{25}{216}}\frac{\theta(y^{-4} q^{17 / 27};q)^{2}}{\theta(y q^{5 / 54};q) \theta(y^{2} q^{5 / 27};q)}~.
\eea
where \(\theta(x ; q)=\prod_{i=0}^{\infty}\left(1-x q^{i}\right)\left(1-q^{i+1} / x\right)\), and $J_0$ is the $\U(1)_\ell$ charge. Note that $q = e^{2 \pi i \tau}$, $y = e^{2 \pi i z}$.

Among chiral primary states $(\overline{L}_{0}=\overline{J}_{0}/2)$ in the right-moving sector that contribute to the elliptic genus, the state subject to \(L_{0}=\mathfrak{q} / 2\) in the left-moving sector form the topological heterotic ring $\cH_{\textrm{top}}$ \cite{Katz:2004nn,Adams:2005tc} where \(\mathfrak{q}\) is equal to the $\U(1)_\cR$ charge \(r_{\phi}\) for a chiral field and \(r_{\psi}-1\) for a Fermi field.
Since the numbers of chiral and Fermi multiplets are equal in the LG theory, it receives contributions only from chiral multiplets with $L_0=\overline J_0/2$, which is isomorphic to the Jacobi ring of the $J$-term
\bea\label{top-het}
\cH_{\textrm{top}}&= \bC[\phi_1,\phi_2]/(\phi_1^4+\phi_2^2,\phi_1^2\phi_2)\cr
&\cong\textrm{Span}[\phi_1^i]_{i=0}^5\oplus \textrm{Span}[\phi_2,\phi_1\phi_2]~.
\eea
In fact, the holomorphic part of the stress-energy tensor \cite{Silverstein:1994ih,Dedushenko:2015opz} is written as
\begin{align}\label{SE-tensor}
T&=\sum_{a=1}^{2}\left[\left(1-\frac{r_{\phi_a}}{2}\right) \partial \phi_{a} \partial \overline{\phi}_{a}-\frac{r_{\phi_a}}{2} \phi_{a} \partial^{2} \overline{\phi}_{a}\right]\\
&+\sum_{a=1}^{2}\left[\frac{i}{2}\left(1+r_{\psi_a}\right) \psi_{a} \partial \overline{\psi}_{a}-\frac{i}{2}\left(1-r_{\psi_a}\right) \partial \psi_{a} \overline{\psi}_{a}\right]~,\nonumber
\end{align}
and  the OPE of a generator of $\cH_{\textrm{top}}$ with the stress-energy tensor shows that it is a primary state with $L_0=\overline J_0/2$.

\section{Modular invariant partition function}
Our goal is to find the modular invariant partition function of the IR CFT in the NS-NS sector as the following form
\begin{align}
Z&= \Tr_{\textrm{NSNS}} q^{L_{0}-\frac{c}{24}} y^{J_{0}}\overline q^{\overline L_{0}-\frac{\overline{c}}{24}} \overline x^{\overline J_{0}}\\
&=\sum_{\textrm{wts}} N^{\SU(2)}_{\ell\overline\ell}N^{\U(1)}_{\nu\lambda\overline{\beta}}~ \chi_{\ell,\nu}^{\PF_{25}}(\tau) \chi_{\lambda}^{\U(1)_{\frac{27}2}}(\tau,z) \cdot \overline{\chi}_{\overline{\ell},\overline \beta}^{\MM_{25}}(\overline\tau,\overline w)\nonumber
\end{align}
which is consistent with the elliptic genus \eqref{EG}, where $\overline q= e^{-2 \pi i \overline \tau}$, $\overline x= e^{-2 \pi i \overline w}$ and `wts' stands for all the weights labeled by $l,\overline{l}, \nu, \lambda, \overline{\beta}$. To obtain an elliptic genus from a partition function, we fix the right-moving sector to be chiral primary states $(\overline{L}_{0}=\overline{J}_{0}/2)$ only from which the elliptic genus receives contributions. Then, we insert $(-1)^F$ or equivalently $(-1)^{2(L_0-\overline L_0)}$ in each term of the left-moving sector \cite{Gadde:2014ppa}.

For this purpose, we shall find the modular invariant combination of $\U(1)$ WZNW characters by following \cite{Gannon:1996hp,Gadde:2016khg}.
In fact, the quadratic forms given by $\U(1)$ levels are rationally equivalent
$$
\textrm{diag}\left (\frac{27}2,27\right)=R^T\, \textrm{diag}(25,2) \,R
$$
where
$$
R=\frac{1}{10}\left(\begin{array}{ll}{2} & {10} \\ {25} & {-10}\end{array}\right)~.
$$
This rational equivalence gives rise to an identity among theta functions
\bea
&\chi^{\U(1)_{25}}_{\mu}(\tau,2 u+10 v)(\chi^{\U(1)_2}_{0}+\chi^{\U(1)_2}_{2})(\tau,25 u - 10v)\cr
&=\sum_{i=0}^{27\times 10-1}\left\{ \chi^{\U(1)_\frac{27\times 10^2}{2}}_{2 \mu+50 i}(\tau,u) \chi^{\U(1)_{27\times 10^2}}_{10 \mu-20 i}(\tau,v)\right.\cr
&\hspace{2cm}+ \left. \chi^{\U(1))_\frac{27\times 10^2}{2}}_{2 \mu+50 i}(\tau,u) \chi^{\U(1)_{27\times 10^2}}_{10 \mu-20 i+27\times 10^2}(\tau,v)\right\}\cr
&\equiv \sum_{\lambda',\rho'} A_{\mu\lambda'{\rho'}}~
\chi^{\U(1)_{\frac{27\times 10^2}{2}}}_{\lambda'}(\tau,u)~\chi^{\U(1)_{27\times 10^2}}_{\rho'}(\tau,v)~.\nonumber
\eea
Furthermore, there is another identity of theta functions
\bea
&\chi^{\U(1)_{\frac{27}2}}_{{\lambda}}(\tau,10u)~\chi^{\U(1)_{27}}_{{\rho}}(\tau,10v)\cr
&=\sum_{i_1,i_2\in \mathbb{Z}_{10}} \chi^{\U(1)_{\frac{27\times 10^2}{2}}}_{10({\lambda}+27 i_1)}(\tau,u)~
\chi^{\U(1)_{27\times 10^2}}_{10({\rho}+54i_2)}(\tau,v) \\
&\equiv \sum_{\lambda',\rho'}B_{{\lambda}{\rho}\lambda'\rho'}~ \chi^{\U(1)_{\frac{27\times 10^2}{2}}}_{{\lambda'}}(\tau,u)~
\chi^{\U(1)_{27\times 10^2}}_{{\rho'}}(\tau,v)~.\nonumber
\eea
From these identities, one can construct the $\U(1)$ modular invariant tensor by
\be\nonumber
N^{\U(1)}_{\nu{\lambda}\overline{\beta}}=\sum_{\lambda^{\prime}, \beta^{\prime}} A_{\nu, {\lambda}^{\prime}, {\beta}^{\prime}} B_{{\lambda}, \overline{\beta}, \lambda^{\prime}, \beta^{\prime}}~,
\ee
which satisfies
$$(M_{\nu'\nu}^{\U(1)_{25}})^* M_{\lambda'\lambda}^{\U(1)_{27/2}} N^{\U(1)}_{\nu\lambda\overline{\beta}} M_{\overline{\beta}\overline{\beta}'}^{\U(1)_{27}} =N^{\U(1)}_{\nu'\lambda'\overline{\beta}'}~,$$
for all $M\in \SL(2,\bZ)$.
More explicitly, one can write
\bea\label{PF}
Z=\sum_{\textrm{wts}} N^{\SU(2)}_{\ell\overline\ell} \chi_{\ell,5m}^{\PF_{25}}(\tau) \chi_{\frac{27m-5n}{2}}^{\U(1)_{\frac{27}2}}(\tau,z) \cdot \overline{\chi}_{\overline   \ell,n}^{\MM_{25}}(\overline\tau,\overline w)
\eea
where the summation over weights runs $m\in \bZ_{10}$, $n\in\bZ_{54}$ and $\ell, \overline\ell \in \bZ_{26}$.

Next, we need to determine the $\SU(2)$ modular invariant tensor $N^{\SU(2)}_{\ell\overline\ell}$.
For the $\SU(2)$ level $k=25$, only the diagonal (type-A) combination $N^{\SU(2)}_{\ell\overline\ell} =\delta_{\ell\overline\ell}$ is listed in the ADE classification \cite{cappelli1987modular,cappelli1987ade,kato1987classification,Gannon:1999cp}. However, with the diagonal $\SU(2)$ combination, one can check that
the partition function would not realize the elliptic genus \eqref{EG}.

To circumvent this situation, we need to relax some of the assumptions in \cite{cappelli1987modular,cappelli1987ade,kato1987classification,Gannon:1999cp} for the classification.
We notice that the following matrix commutes with all the modular matrices $M^{\SU(2)}$
\bea\nonumber
N^{\textrm{nd}}_{\ell\overline \ell}=&(\delta_{2,\ell}-\delta_{14,\ell}+\delta_{20,\ell})(\delta_{2,\overline\ell}-\delta_{14,\overline\ell}+\delta_{20,\overline\ell})\cr
&+(\delta_{5,\ell}-\delta_{11,\ell}+\delta_{23,\ell})(\delta_{5,\overline\ell}-\delta_{11,\overline\ell}+\delta_{23,\overline\ell})
\eea
where the indices range $\ell,\overline \ell \in \bZ_{26}$. Then, we set
\be\label{SU2-comb}
N^{\SU(2)}_{\ell\overline\ell} =\delta_{\ell\overline\ell}-\frac13N^{\textrm{nd}}_{\ell\overline \ell}~.
\ee
This clearly violates the assumption that $N_{i\overline i}$ in \eqref{RCFT-Hilbert} are non-negative integer multiplicities, which has been adopted in the literature including \cite{cappelli1987modular,cappelli1987ade,kato1987classification,Gannon:1999cp}. However, if we use \eqref{SU2-comb} in \eqref{PF}, the partition function is a formal series of $(q,y,\overline q,\overline x)$ with non-negative integer coefficients and it is moreover consistent with the elliptic genus \eqref{EG}. We claim that it is the partition function of the IR CFT.

\section{Hilbert space and a new CFT $\wt \PF_{25}$}
To demystify the multiplicities \eqref{SU2-comb} with negative fractional numbers, let us investigate the Hilbert space of the IR CFT. To this end, we denote by $V_{\ell,m}^{\PF_{25}}$ a highest weight representation of $\PF_{25}$. In addition, by taking the direct sum of $s=0$ and $s=2$ weight of $\U(1)_2$, we write by $V_{\ell,m}^{\MM_{25}}$ a highest weight representation of $\MM_{25}$ in the NS sector.
There are isomorphisms of irreducible modules
\bea\nonumber
& V_{\ell,m}^{\PF_{25}}\cong V_{\ell,50-m}^{\PF_{25}}\cong  V_{25-\ell,m+25}^{\PF_{25}} \cr
& V_{\ell,m}^{\MM_{25}}\cong  V_{\ell,54-m}^{\MM_{25}}\cong  V_{25-\ell,m+27}^{\MM_{25}}~.
\eea

First, we note an identity of the parafermion characters
\bea\label{PF-id}
3&=\sum_{m=0}^4\chi^{\PF_{25}}_{2,10m}-\chi^{\PF_{25}}_{14,10m}+\chi^{\PF_{25}}_{20,10m}\cr
&=\sum_{m=0}^4\chi^{\PF_{25}}_{5,10m+5}-\chi^{\PF_{25}}_{11,10m+5}+\chi^{\PF_{25}}_{23,10m+5}~,
\eea
which counts the number of primary states $|\ell,m\rangle_{\PF_{25}}$ with conformal dimension $h^{\PF_{25}}_{\ell,m}=2/27$ in $\PF_{25}$:
\bea\label{PF-primary}
|\ell,m\rangle_{\PF_{25}}&=|2,0\rangle, ~|20,20\rangle, ~|20,30\rangle, \quad \textrm{or}\cr
|\ell,m\rangle_{\PF_{25}}&=|23,25\rangle, ~|5,5\rangle,~|5,45\rangle~.
\eea
Hence, roughly speaking, the non-diagonal part of \eqref{SU2-comb} adds or eliminates a certain linear combination of these states to or from $V_{\ell,m}^{\PF_{25}}$.

To see how the spectrum is organized, we compare the diagonal spectrum
\bea\label{diag}
\cH_{\textrm{diag}}=\bigoplus_{\ell,m,n}  V_{\ell,5m}^{\PF_{25}}\otimes V_{\frac{27m-5n}{2}}^{\U(1)_{\frac{27}2}} \otimes \overline{V}_{  \ell,n}^{\MM_{25}}~
\eea
where $N^{\SU(2)}_{\ell\overline\ell}=\delta_{\ell\overline\ell}$
with the information of the Hilbert space of the IR CFT obtained from the LG model.
Note that the diagonal spectrum $\cH_{\textrm{diag}}$ contains primary states of $\PF_{25}\times \U(1)_{\frac{27}{2}}\times \MM_{25}$
\bea\label{diag-primaries}
|5s,5s\rangle_{\PF_{25}}\otimes |-s\rangle_{\U(1)_{\frac{27}2}}\otimes |5s,-5s\rangle_{\MM_{25}}~,\cr
|5s,50-5s\rangle_{\PF_{25}}\otimes |-s\rangle_{\U(1)_{\frac{27}2}}\otimes |5s,-5s\rangle_{\MM_{25}}~,
\eea
which obey the condition $L_0= \overline J_0/2=\overline L_0$. Here we have $s=0,1,\ldots,5$ and the states in the first and second line are identical to the vacuum state when $s=0$. Thus, there are ten primary states subject to the condition in $\cH_{\textrm{diag}}$ whereas the topological heterotic ring \eqref{top-het} of the IR CFT is eight-dimensional as a vector space.

Hence, the diagonal spectrum \eqref{diag} is not the actual Hilbert space. To realize the ring structure of $\cH_{\textrm{top}}$ in \eqref{top-het}, let us suppose that $\phi_1$ and $\phi_2$ in $\cH_{\textrm{top}}$ respectively correspond to
\bea\label{identification}
&|5,5\rangle_{\PF_{25}}+|5,45\rangle_{\PF_{25}}\quad \textrm{and}\cr
&|10,10\rangle_{\PF_{25}}-|10,40\rangle_{\PF_{25}}~.
\eea
Here and in what follows, we suppress the parts of $\U(1)_{27/2}$ and $\MM_{25}$ of \eqref{diag-primaries}. Then the fusion rule tells us that $\phi_1^2\phi_2$ corresponds to
\be\label{eliminate1}
|20,20\rangle_{\PF_{25}}-|20,30\rangle_{\PF_{25}}~
\ee
in \eqref{diag-primaries}, which is decoupled from the spectrum due to the equation of motion.
 In addition, there is no generator in $\cH_{\textrm{top}}$ corresponding to
 \be\label{eliminate2}
 |5,5\rangle_{\PF_{25}}-|5,45\rangle_{\PF_{25}}~
 \ee
in \eqref{diag-primaries}. Thus, the IR CFT excludes these two states, \eqref{eliminate1} and \eqref{eliminate2}, and
the identification of \eqref{identification} with $\phi_1$ and $\phi_2$ as well as the fusion rule indeed reproduces the topological heterotic ring  \eqref{top-het}.

On the other hand, one can show that $\phi_1^2\partial\phi_2\sim-2\phi_1\phi_2\partial \phi_1$ is a primary in the IR CFT from the OPE with the stress-energy tensor \eqref{SE-tensor} up to the equations of motion ($\partial W/\partial \phi_i=0=\partial W/\partial \psi_i$ and their complex conjugates). Moreover, $\phi_1^2\partial\phi_2$ and its descendants contribute to the elliptic genus by
\bea\nonumber
&(\chi^{\PF_{25}}_{|20,20>-|20,30>} -1)\chi^{\U(1)_{27/2}}_{-4}=(\chi^{\PF_{25}}_{20,20} -1)\chi^{\U(1)_{27/2}}_{-4} \cr
&=(q+3 q^2+6 q^3+12 q^4+21 q^5+\cdots) \chi^{\U(1)_{27/2}}_{-4}~.
\eea
Ignoring the $\U(1)_{27/2}$ part, the subtraction by one means the omission of \eqref{eliminate1}, and the primary $\phi_1^2\partial\phi_2$ contributes to $q^{1}$ whereas the subsequent higher order terms count its descendants.
This implies that although the IR CFT is not endowed with the parafermionic symmetry $\SU(2)_{25}/\U(1)_{25}$,  it is still a character of a module of the Virasoro algebra in the left-moving sector.
Similarly, it is easy to check from the OPE that $\phi_1^3\partial \overline\phi_2 \sim -2\phi_2\partial \overline\phi_1$ is also a primary in the IR CFT, and the contribution from its conformal family to the elliptic genus is $(\chi^{\PF_{25}}_{|5,5>-|5,45>} -1)\chi^{\U(1)_{27/2}}_{-1}$.

Furthermore, an explicit computation using \eqref{PF-id} shows that the elliptic genus \eqref{EG} receives all the contributions from the part of $\ell=5,n=-5$ in $\cH_{\textrm{diag}}$
except the states \eqref{eliminate1} and \eqref{eliminate2}, and their $\U(1)_{27/2}$ descendants.
Indeed, the Hilbert space is organized at the IR fixed point in such a way that the states \eqref{eliminate1} and \eqref{eliminate2} are excluded in the $\PF_{25}$ part but it preserves the Virasoro symmetry and the modular invariance. Denoting the CFT of central charge $16/9$ by $\wt \PF_{25}$, the $\ell=5,20$ parts of the Hilbert space are isomorphic to the quotient spaces
\bea\nonumber
\cH_{5}^{\wt\PF_{25}}&\cong \bigoplus_{m=0}^{4} V_{5,10m+5}^{\PF_{25}}\Big/\bC ( |5,5\rangle-|5,45\rangle)~,\cr
\cH_{20}^{\wt\PF_{25}}&\cong \bigoplus_{m=0}^{4} V_{20,10m}^{\PF_{25}}\Big/\bC(|20,20\rangle-|20,30\rangle)~,
\eea
as vector spaces graded by $L_0$.

In order to keep the modular invariance, one needs to arrange the primary states \eqref{PF-primary} of $\PF_{25}$ according to the non-diagonal part of \eqref{SU2-comb}:
\bea\label{wt-PF}
\cH_{2}^{\wt\PF_{25}}&\cong\bigoplus_{m=0}^{4} V_{2,10m}^{\PF_{25}}\Big/\bC |2,0 \rangle~,\cr
\cH_{23}^{\wt\PF_{25}}&\cong\bigoplus_{m=0}^{4} V_{23,10m+5}^{\PF_{25}}\Big/\bC |23,25 \rangle~,\cr
\cH_{14}^{\wt\PF_{25}}&\cong \bC (|20,20\rangle+|20,30\rangle)\oplus \bigoplus_{m=0}^{4}  V_{14,10m}^{\PF_{25}}~,\cr
\cH_{11}^{\wt\PF_{25}}&\cong  \bC ( |5,5\rangle+|5,45\rangle)\oplus \bigoplus_{m=0}^{4}  V_{11,10m+5}^{\PF_{25}}~.
\eea
Here, $\cong$ means an isomorphism as vector spaces graded by $L_0$. For the other $\ell\neq 2,5,11,14,20,23$, they are isomorphic to those of $\PF_{25}$ $$\cH_{\ell}^{\wt \PF_{25}}\cong\bigoplus_{m=0}^{4} V_{\ell,10m+5(\ell~\mathrm{mod}~2)}^{\PF_{25}}~.$$

All in all, the Hilbert space of the IR CFT is then expressed as
\be\label{Hilbert-sp}
\cH=\bigoplus_{\ell,n} \cH_{\ell}^{\wt \PF_{25}}\otimes V_{\frac{27\ell-5n}{2}}^{\U(1)_{\frac{27}2}} \otimes  \overline{V}_{\ell,n}^{\MM_{25}}~,
\ee
whose generating function is \eqref{PF} with \eqref{SU2-comb}. This explains the reason why the partition function \eqref{PF} with \eqref{SU2-comb} is a formal power series with non-negative integer coefficients.
If we restrict the right-moving sector to be chiral primary states, we have
$$
\cH\big|_{\overline L_0=\overline J_0/2}=\bigoplus_{\ell} \cH_{\ell}^{\wt \PF_{25}}\otimes V_{16\ell}^{\U(1)_{\frac{27}2}}~,
$$
which exactly reproduces the elliptic genus \eqref{EG} by appropriately including signs. In fact, under the equations of motion, the conformal families of two primaries  $\psi_{i}\partial\overline\phi_1$ ($i=1,2$) contribute to \eqref{EG}
\bea\label{2-primaries}
&(\chi^{\PF_{25}}_{2,0} -1)\chi^{\U(1)_{27/2}}_{5}\\
&=(2 q+3 q^2+6 q^3+10 q^4+18 q^5+\cdots) \chi^{\U(1)_{27/2}}_{5}~.
\eea
For $\ell=23$,  those of two primaries $\overline \psi_i\phi_1^2(\partial\phi_2)^2$  ($i=1,2$) yield the contribution $(\chi^{\PF_{25}}_{23,25} -1)\chi^{\U(1)_{27/2}}_{17}$. In addition, the conformal family of a primary $\psi_1\psi_2$ combines the two irreducible characters of $\PF_{25}$ into one ``irreducible'' character of $\wt\PF_{25}$
\bea
&(1+\chi^{\PF_{25}}_{14,20}+\chi^{\PF_{25}}_{14,30})\chi^{\U(1)_{27/2}}_{8}\cr
&=(1+2 q + 4 q^2 + 10 q^3 + 20 q^4 + 38 q^5+\cdots) \chi^{\U(1)_{27/2}}_{8}.\nonumber
\eea
In a similar fashion, that of a primary $\overline \psi_1\overline \psi_2 \phi_1\phi_2(\partial \phi_1)^2$ gives the contribution $(1+\chi^{\PF_{25}}_{11,5}+\chi^{\PF_{25}}_{11,45})\chi^{\U(1)_{27/2}}_{14}$. Hence, this provides a strong evidence that the graded vector spaces \eqref{wt-PF} are decomposed into modules of the Virasoro algebra and $\wt \PF_{25}$ preserves the conformal symmetry.
In conclusion, the $\cN=(0,2)$ LG model flows to
$$
\left(\wt \PF_{25}\times \U(1)_{\frac{27}2}\right)  \otimes \overline{\left( \frac{\SU(2)_{25}\times \U(1)_2}{\U(1)_{27}} \right)}~,
$$
and the modular invariant Hilbert space  \eqref{Hilbert-sp} on a torus is decomposed into modules of the left-moving Virasoro algebra and the right-moving $\cN=2$ super-Virasoro algebra.

\section{Discussions}
We find the modular invariant partition function beyond the ADE classification \cite{cappelli1987modular,cappelli1987ade,kato1987classification,Gannon:1999cp} because a part of the left-moving sector is the new CFT $\wt \PF_{25}$ obtained by breaking the parafermionic symmetry of $\PF_{25}$. Certainly, more investigation needs to be carried out to understand $\wt \PF_{25}$. In particular, it is desirable to determine two ``irreducible'' characters of the primaries $\psi_{i}\partial\overline\phi_1$ (resp.  $\overline \psi_i\phi_1^2(\partial\phi_2)^2$) in $\wt \PF_{25}$ whose sum is equal to $\chi^{\PF_{25}}_{2,0} -1$ in \eqref{2-primaries} (resp. $\chi^{\PF_{25}}_{23,25} -1$).

In \cite{Gholson:2018zrl}, $\cN=(0,2)$ LG models with the same left and right central charges $\le 3$ have been classified. In the classification of \cite{Gholson:2018zrl}, IR CFTs of Class 2.a with superpotential
$$
\psi_1(\phi_1^m+\phi_2^n)+\psi_2\phi_1\phi_2~, \quad m,n\in \bZ_{>0}~,
$$
are described by diagonal modular pairing of PFs and $\U(1)$ WZNW models in the left-moving-sector and $\cN=2$ MMs in the right-moving sector  \cite{Gadde:2016khg}. This is because their topological heterotic rings are simple and it does not contain a mixed generator like $\phi_1\phi_2$. Like in our example, the topological heterotic rings of the other classes in \cite{Gholson:2018zrl} are more complicated, and we observe that their elliptic genera cannot be realized by characters of PFs and $\U(1)$ WZNW models except our example \eqref{potential}. (Another exception is Class 2.b with $k=3$, but it is equivalent to $\cN=(2,2)$ MM of type $E_7$.) It is expected that the left-moving sectors of IR CFTs would be unknown ones so that it requires further study to understand how $J$-terms of $\cN=(0,2)$ LG models are encoded in IR CFTs. It is also worth mentioning that the condition of the same left and right-moving central charges in \cite{Gholson:2018zrl} is rather special in $\cN=(0,2)$ LG models, and a vast class of general $\cN=(0,2)$ LG models are waiting to be investigated.

Since A.B. Zamolodchikov has identified the LG/CFT correspondence \cite{zamolodchikov1986conformal}, it has given drastically new insights in quantum field theories and mathematical physics. This article just takes a peek at the LG/CFT correspondence with $\cN=(0,2)$ supersymmetry, but we hope that our example shows its fertility and will intensify further study on it.

\section{Acknowledgments}
S.N. has first come across 2d CFT through a little gem \cite{Eguchi-Yang} presented by T.Eguchi and S.-K.Yang. Their works including articles and textbooks in Japanese have been an inspiration to him. This paper intersects with their interests so that it is dedicated to their academic achievements with admiration and gratitude.
We thank D.Yokoyama for collaboration at the initial stage of this project. We are grateful to I.Gahramanov, Y.Nakayama, D.Pei, P.Putrov, D.Roggenkamp and E.Sharpe for discussion and correspondence, and especially to M.Dedushenko for sending a private note based on \cite{Dedushenko:2015opz} for a consistency check of chiral primary states. S.N. is indebted to Fudan University for providing him an opportunity to teach 2d CFT during which he came up with the idea of this project. S.N. also thanks IHES, Mittag Leffler Institute, and QGM at Aarhus University for warm hospitality where a part of the work was carried out. We acknowledge the support from NSFC Grant No.11850410428.

\appendix

\section{Notations}
Here, we summarize convention and definitions necessary in this article.
$\U(1)_k$ and  $\SU(2)_k$  characters are given by
\begin{align}
\chi_{m}^{\U(1)_k}(\tau, z)&=\frac{\Theta_{m,k}(\tau, z)}{\eta(\tau)}~,\cr
\chi_{\ell}^{\SU(2)_k}(\tau, z)&=\frac{\Theta_{\ell+1, k+2}(\tau, z)-\Theta_{-(\ell+1), k+2}(\tau, z)}{\Theta_{1,2}(\tau, z)-\Theta_{-1,2}(\tau, z)}~,\nonumber
\end{align}
where $\eta(\tau)=q^{\frac{1}{24}}\prod_{m=1}^\infty (1-q^m)$ is the Dedekind eta-function and the theta function is defined as
\begin{equation}\nonumber
\Theta_{m, k}(\tau, z) \equiv \sum_{n \in \mathbb{Z}} q^{k\left(n+\frac{m}{2 k}\right)^{2}} y^{k\left(n+\frac{m}{2 k}\right)}~.
\end{equation}
The weights of $\U(1)_k$ and $\SU(2)_k$ run over $m=0,\ldots,2k-1$ and $\ell=0,\ldots,k$, respectively. It is well-known that the modular group $\SL(2,\bZ)$ is generated by $T$ and $S$, and a $T$-transformation on characters of a chiral algebra $\cA$ is always diagonalizable
$$ \chi_r^\cA(\tau+1)=e^{2\pi i (h_r-c/24)}\chi_r^\cA(\tau)~,$$
where $h_r$ is the conformal dimension of the corresponding highest weight state.
Under the $S$-transformation, the characters of $\cA_k=\U(1)_k,\SU(2)_k$ are transformed as
\begin{align}\nonumber
\chi_{r}^{\cA_k}\left(-\frac{1}{\tau},\frac{z}{\tau}\right)&=e^{\frac{i\pi kz^2}{2\tau}} \sum_{r'} S_{rr'}^{\cA_k}\chi_{r'}^{\cA_k}(\tau, z)~,
\end{align}
where
\begin{align}\nonumber
&S_{\ell, \ell^{\prime}}^{\SU(2)_k} \equiv \sqrt{\frac{2}{k+2}} \sin \left(\frac{\pi(\ell+1)\left(\ell^{\prime}+1\right)}{k+2}\right),\cr
& S_{m, m^{\prime}}^{\U(1)_k} \equiv \frac{1}{\sqrt{2 k}} e^{-2 \pi i \frac{m m^{\prime}}{2 k}}~.
\end{align}

A character of a coset model $\cA/\cB$ can be computed via a branching rule
$$
V_{\ell}^\cA =\bigoplus_m V_{m}^\cB \oplus V_{\ell,m}^{\cA/\cB}
$$
where $V_{\ell}^\cA$ and $V_{m}^\cB$ are highest weight representations of the chiral algebra $\cA$ and $\cB$, respectively.
By defining the string function $c_{\ell, m}^{(k)}$ \cite{Gepner:1986hr}
$$
\chi_{\ell}^{\SU(2)_k}(\tau, z)=\sum_{m \in \bZ_{2 k}} c_{\ell, m}^{(k)}(\tau) \Theta_{m, k}(\tau, z)~,
$$
a character of the parafermion is then expressed as
\begin{equation}\nonumber
\chi_{\ell, m}^{\PF_k}(\tau)=\eta(\tau) c_{\ell, m}^{(k)}(\tau)~,
\end{equation}
where $\ell+m\in 2\bZ$, and otherwise $\chi_{\ell,m}^{\PF_k}=0$. Note that the characters obey $\chi_{\ell, m}^{\PF_k}=\chi_{\ell,2k- m}^{\PF_k}=\chi_{k-\ell, m+k}^{\PF_k}$.

In addition, a character of the $\cN=2$ minimal model in the NS sector \cite{Ravanini:1987yg} is given by
$$
\chi_{\ell,m}^{\MM_k}(\tau, z)=\sum_{r \in \bZ_{2k}} c_{\ell, r}^{(k)}(\tau) \Theta_{(k+2)r-km, k(k+2)}\left(\frac{\tau}2, \frac{z}{k+2}\right)
$$
where the weights $s=0,2$ of $\U(1)_2$ are summed. Note that the weights are subject to $\ell+m\in 2\bZ$, and otherwise $\chi_{\ell,m}^{\MM_k}=0$. Note that the characters satisfy $\chi_{\ell, m}^{\MM_k}=\chi_{\ell, 2(k+2)-m}^{\MM_k}=\chi_{k-\ell, m+k+2}^{\MM_k}$.

\bibliographystyle{ytphys}

\bibliography{conformal-ref}

\providecommand{\href}[2]{#2}\begingroup\raggedright\begin{thebibliography}{10}

\bibitem{belavin1984infinite}
A.~A. Belavin, A.~M. Polyakov, and A.~B. Zamolodchikov, ``{Infinite Conformal
  Symmetry in Two-Dimensional Quantum Field Theory},''
\href{http://dx.doi.org/10.1016/0550-3213(84)90052-X}{{\em Nucl. Phys.}
  {\bfseries B241} (1984) 333--380}.
%%CITATION = NUPHA,B241,333;%%.

\bibitem{cardy1986operator}
J.~L. Cardy, ``{Operator Content of Two-Dimensional Conformally Invariant
  Theories},''
\href{http://dx.doi.org/10.1016/0550-3213(86)90552-3}{{\em Nucl. Phys.}
  {\bfseries B270} (1986) 186--204}.
%%CITATION = NUPHA,B270,186;%%.

\bibitem{itzykson1988conformal}
C.~Itzykson, H.~Saleur, and J.-B. Zuber, {\em {Conformal invariance and
  applications to statistical mechanics}}.
\newblock World Scientific, 1988.

\bibitem{Kac:1984mq}
V.~G. Kac and D.~H. Peterson, ``{Infinite dimensional Lie algebras, theta
  functions and modular forms},''
\href{http://dx.doi.org/10.1016/0001-8708(84)90032-X}{{\em Adv. Math.}
  {\bfseries 53} (1984) 125--264}.
%%CITATION = ADMTA,53,125;%%.

\bibitem{cappelli1987modular}
A.~Cappelli, C.~Itzykson, and J.~B. Zuber, ``{Modular Invariant Partition
  Functions in Two-Dimensions},''
\href{http://dx.doi.org/10.1016/0550-3213(87)90155-6}{{\em Nucl. Phys.}
  {\bfseries B280} (1987) 445--465}.
%%CITATION = NUPHA,B280,445;%%.

\bibitem{cappelli1987ade}
A.~Cappelli, C.~Itzykson, and J.~B. Zuber, ``{The ADE Classification of Minimal
  and A1(1) Conformal Invariant Theories},''
\href{http://dx.doi.org/10.1007/BF01221394}{{\em Commun. Math. Phys.}
  {\bfseries 113} (1987) 1}.
%%CITATION = CMPHA,113,1;%%.

\bibitem{kato1987classification}
A.~Kato, ``{Classification of Modular Invariant Partition Functions in
  Two-dimensions},''
\href{http://dx.doi.org/10.1142/S0217732387000732}{{\em Mod. Phys. Lett.}
  {\bfseries A2} (1987) 585}.
%%CITATION = MPLAE,A2,585;%%.

\bibitem{Gannon:1999cp}
T.~Gannon, ``{The Cappelli-Itzykson-Zuber A-D-E classification},''
  \href{http://dx.doi.org/10.1142/S0129055X00000265}{{\em Rev. Math. Phys.}
  {\bfseries 12} (2000) 739--748},
\href{http://arxiv.org/abs/math/9902064}{{\ttfamily arXiv:math/9902064}}.
%%CITATION = MATH/9902064;%%.

\bibitem{goddard1986unitary}
P.~Goddard, A.~Kent, and D.~Olive, ``{Unitary representations of the Virasoro
  and super-Virasoro algebras},''
  \href{http://dx.doi.org/10.1007/BF01464283}{{\em Commun. in Math. Phys.}
  {\bfseries 103} no.~1, (1986) 105--119}.

\bibitem{Gepner:1986hr}
D.~Gepner and Z.-A. Qiu, ``{Modular Invariant Partition Functions for
  Parafermionic Field Theories},''
\href{http://dx.doi.org/10.1016/0550-3213(87)90348-8}{{\em Nucl. Phys.}
  {\bfseries B285} (1987) 423}.
%%CITATION = NUPHA,B285,423;%%.

\bibitem{Ravanini:1987yg}
F.~Ravanini and S.-K. Yang, ``{Modular Invariance in $N=2$ Superconformal Field
  Theories},''
\href{http://dx.doi.org/10.1016/0370-2693(87)91194-4}{{\em Phys. Lett.}
  {\bfseries B195} (1987) 202--208}.
%%CITATION = PHLTA,B195,202;%%.

\bibitem{Vafa:1988uu}
C.~Vafa and N.~P. Warner, ``{Catastrophes and the Classification of Conformal
  Theories},''
\href{http://dx.doi.org/10.1016/0370-2693(89)90473-5}{{\em Phys. Lett.}
  {\bfseries B218} (1989) 51--58}.
%%CITATION = PHLTA,B218,51;%%.

\bibitem{Martinec:1988zu}
E.~J. Martinec, ``{Algebraic Geometry and Effective Lagrangians},''
\href{http://dx.doi.org/10.1016/0370-2693(89)90074-9}{{\em Phys. Lett.}
  {\bfseries B217} (1989) 431--437}.
%%CITATION = PHLTA,B217,431;%%.

\bibitem{Witten:1993yc}
E.~Witten, ``{Phases of $N=2$ theories in two-dimensions},''
  \href{http://dx.doi.org/10.1016/0550-3213(93)90033-L}{{\em Nucl. Phys.}
  {\bfseries B403} (1993) 159--222},
\href{http://arxiv.org/abs/hep-th/9301042}{{\ttfamily arXiv:hep-th/9301042}}.
%%CITATION = HEP-TH/9301042;%%.

\bibitem{Gadde:2016khg}
A.~Gadde and P.~Putrov, ``{Exact solutions of (0,2) Landau-Ginzburg models},''
\href{http://arxiv.org/abs/1608.07753}{{\ttfamily arXiv:1608.07753 [hep-th]}}.
%%CITATION = ARXIV:1608.07753;%%.

\bibitem{Gholson:2018zrl}
S.~M. Gholson and I.~V. Melnikov, ``{Small Landau-Ginzburg theories},''
  \href{http://dx.doi.org/10.1007/JHEP04(2019)132}{{\em JHEP} {\bfseries 04}
  (2019) 132},
\href{http://arxiv.org/abs/1811.06105}{{\ttfamily arXiv:1811.06105 [hep-th]}}.
%%CITATION = ARXIV:1811.06105;%%.

\bibitem{Benini:2012cz}
F.~Benini and N.~Bobev, ``{Exact two-dimensional superconformal R-symmetry and
  c-extremization},''
  \href{http://dx.doi.org/10.1103/PhysRevLett.110.061601}{{\em Phys. Rev.
  Lett.} {\bfseries 110} no.~6, (2013) 061601},
\href{http://arxiv.org/abs/1211.4030}{{\ttfamily arXiv:1211.4030 [hep-th]}}.
%%CITATION = ARXIV:1211.4030;%%.

\bibitem{Dedushenko:2015opz}
M.~Dedushenko, ``{Chiral algebras in Landau-Ginzburg models},''
  \href{http://dx.doi.org/10.1007/JHEP03(2018)079}{{\em JHEP} {\bfseries 03}
  (2018) 079},
\href{http://arxiv.org/abs/1511.04372}{{\ttfamily arXiv:1511.04372 [hep-th]}}.
%%CITATION = ARXIV:1511.04372;%%.

\bibitem{Katz:2004nn}
S.~H. Katz and E.~Sharpe, ``{Notes on certain (0,2) correlation functions},''
  \href{http://dx.doi.org/10.1007/s00220-005-1443-1}{{\em Commun. Math. Phys.}
  {\bfseries 262} (2006) 611--644},
\href{http://arxiv.org/abs/hep-th/0406226}{{\ttfamily arXiv:hep-th/0406226}}.
%%CITATION = HEP-TH/0406226;%%.

\bibitem{Adams:2005tc}
A.~Adams, J.~Distler, and M.~Ernebjerg, ``{Topological heterotic rings},''
  \href{http://dx.doi.org/10.4310/ATMP.2006.v10.n5.a2}{{\em Adv. Theor. Math.
  Phys.} {\bfseries 10} no.~5, (2006) 657--682},
\href{http://arxiv.org/abs/hep-th/0506263}{{\ttfamily arXiv:hep-th/0506263}}.
%%CITATION = HEP-TH/0506263;%%.

\bibitem{Silverstein:1994ih}
E.~Silverstein and E.~Witten, ``{Global U(1) R symmetry and conformal
  invariance of (0,2) models},''
  \href{http://dx.doi.org/10.1016/0370-2693(94)91484-2}{{\em Phys. Lett.}
  {\bfseries B328} (1994) 307--311},
\href{http://arxiv.org/abs/hep-th/9403054}{{\ttfamily arXiv:hep-th/9403054}}.
%%CITATION = HEP-TH/9403054;%%.

\bibitem{Gadde:2014ppa}
A.~Gadde, S.~Gukov, and P.~Putrov, ``{Exact Solutions of 2d Supersymmetric
  Gauge Theories},'' \href{http://dx.doi.org/10.1007/JHEP11(2019)174}{{\em
  JHEP} {\bfseries 11} (2019) 174},
\href{http://arxiv.org/abs/1404.5314}{{\ttfamily arXiv:1404.5314 [hep-th]}}.
%%CITATION = ARXIV:1404.5314;%%.

\bibitem{Gannon:1996hp}
T.~Gannon, ``{$U(1)^m$ modular invariants, N=2 minimal models, and the quantum
  Hall effect},'' \href{http://dx.doi.org/10.1016/S0550-3213(97)00032-1}{{\em
  Nucl. Phys.} {\bfseries B491} (1997) 659--688},
\href{http://arxiv.org/abs/hep-th/9608063}{{\ttfamily arXiv:hep-th/9608063}}.
%%CITATION = HEP-TH/9608063;%%.

\bibitem{zamolodchikov1986conformal}
A.~B. Zamolodchikov, ``{Conformal Symmetry and Multicritical Points in
  Two-Dimensional Quantum Field Theory},''
{\em Sov. J. Nucl. Phys.} {\bfseries 44} (1986) 529--533.
%%CITATION = SJNCA,44,529;%%.

\bibitem{Eguchi-Yang}
T.~Eguchi and S.-K. Yang,
  ``{\href{https://www.jstage.jst.go.jp/article/butsuri1946/44/12/44_12_894/_pdf/-char/ja}{Virasoro
  algebra and critical phenomena.} (In Japanese)},''
  \href{http://dx.doi.org/10.11316/butsuri1946.44.894}{{\em Journal of the
  Physical Society of Japan} {\bfseries 44} no.~12, (1989) 894--901}.

\end{thebibliography}\endgroup

\end{document}